\documentclass[aps,12pt,a4paper]{revtex4-2}
\usepackage{epsfig}
\usepackage{graphicx}
\usepackage{amsmath,amssymb,color}
\usepackage[english]{babel}
\usepackage{natbib}
\usepackage{graphicx}
\usepackage{float}
\parskip=\medskipamount
\usepackage{subfigure}

\newcommand{\eq}[1]{(\ref{#1})}
\newcommand{\fig}[1]{Fig.~\ref{#1}}

\newcommand{\be}{\begin{equation}}
\newcommand{\ee}{\end{equation}}
\newcommand{\beq}{\begin{equation}}
\newcommand{\eeq}{\end{equation}}
\newcommand\disp{\displaystyle}

\newcommand{\la}{\left<}
\newcommand{\ra}{\right>}

\begin{document}

\title{Unconventional critical behavior of polymers at sticky boundaries}

\author{Alexander Gorsky}
\email{shuragor@mail.ru}
\affiliation{Institute for Information Transmission Problems RAS, 127051 Moscow, Russia}
\affiliation{Laboratory of Complex Networks, Center for Neurophysics and Neuromorphic Technologies}
\author{Sergei Nechaev}
\email{sergei.nechaev@gmail.com }
\affiliation{LPTMS, CNRS -- Universit\'e Paris Saclay, 91405 Orsay Cedex, France}
\affiliation{Laboratory of Complex Networks, Center for Neurophysics and Neuromorphic Technologies}
\author{Alexander Valov}
\email{aleksandr.valov@mail.huji.ac.il}
\affiliation{Racah Institute of Physics, Hebrew University of Jerusalem, Jerusalem 91904, Israel}


\begin{abstract}
We discuss the generalization of a classical problem involving an $N$-step ideal polymer adsorption at a sticky boundary (potential well of depth $U$). It is known that as $N$ approaches infinity, the path undergoes a 2nd-order localization transition at a certain value of $U_{\text{tr}}$. By considering the random walk on a half-line with a sticky boundary (Model I), we demonstrate that the order of the phase transition can be altered by adjusting the scaling of the first return probability to the boundary. Additionally, we present a model of a random path on a discrete 1D lattice with non-uniform local hopping amplitudes and a potential well at the boundary (Model II). We illustrate that one can tailor such amplitudes so that the polymer undergoes a 3rd-order phase transition.
\end{abstract}

\maketitle

\section{Introduction}
\label{sect:1}

Despite having a long history, the issue of polymer adsorption continues to be a prominent topic in modern statistical physics of polymers. The sustained interest in the adsorption and localization of polymer chains at surfaces and interfaces serves a dual purpose. On one hand, this interest revolves around establishing a fundamental connection between polymer statistics and a general theory of phase transitions in condensed matter physics. On the other hand, it is driven by the practical applications in polymer chemistry and the design of new tailor-made materials.

Without claiming to be exhaustive, let us outline the main directions of polymer physics that have evolved from various aspects of polymer adsorption. Primarily, there is the coil-to-globule phase transition problem rooted in the localization of a random walk at the potential well of depth $U$ in three-dimensional space (see \cite{borisov1, borisov2} and references therein). Understanding the fundamental principles behind polymer adsorption on sticky surfaces has proven crucial in the investigation of polymers grafted at surfaces \cite{brush} (polymer brushes). Another class of problems originating from random walk adsorption at potential wells deals with localization at interfaces \cite{hollander}, inhomogeneities in space \cite{sommer}, or networks (graphs) \cite{nyrkova}. The potential well in these cases often has an entropic nature. A specific class of such problems has been studied in \cite{heavy, heavy2}, where the localization of random walks on regular graphs (trees) with a single special vertex (called a ``heavy root''), decorated stars, and hyperbolic graphs have been analyzed. In all those cases, it has been shown that the presence of a ``heavy root'' may lead to a localization transition if the functionality of a special vertex exceeds some critical value.

In this study, we focus on the following problem. Consider a random path in one-dimensional semi-infinite space with a potential well located at the boundary. Since the boundary is sticky, the system undergoes a localization transition at some critical stickiness $U_{tr}$. Could we control the order of the phase transition by changing the details of a walker's bulk dynamics? Two models (I and II) are under our attention:
\begin{itemize}
\item[I.] The $N$-step random walk on a continuous semi-infinite line $x \in [0, \infty)$ with a specific scaling of the first return probability $P(N) \propto N^{-\alpha}$ ($\alpha > 0$) interacting with a potential well of depth $U$ located at the boundary $x=0$;
\item[II.] The $N$-step inhomogeneous random walk on a discrete lattice $k=1,2,...K$ with the specific scaling of position-dependent hopping amplitudes, $b(k)\propto (K-k)^{\chi}$ ($\chi > 0$), where $k \in [1, K]$ ($K \to \infty$), interacting with a potential well of depth $U$ located at the boundary $k=1$.
We demonstrate that in model I, one can obtain any order $\theta > 0$ of a phase transition by tuning $\alpha$ ($\alpha > 1$). In model II, one can change the 2nd-order phase transition to the 3rd-order one by replacing $\chi=0$ with $\chi=\frac{1}{2}$. Model II can also be visualized as a Brownian bridge (the closed random loop) on a tree (nonhomogeneous in general) with the sticky root at which the potential well is located.
\end{itemize}

The paper is structured as follows. In Section \ref{sect:2}, we revisit the adsorption of the random path at the potential well and pay attention to the dependence of the phase transition order on the return probability. In Section \ref{sect:3}, we study the critical behavior of the walker on the semi-infinite lattice with inhomogeneous hopping rates and analyze the dependence of the transition order on the scaling of hopping amplitudes. Section \ref{sect:4} summarizes the results obtained in the work.

\section{Adsorption of a fractal polymer: First return probability and the phase transition order}
\label{sect:2}

\begin{figure}[ht]
\includegraphics[width=0.8\textwidth]{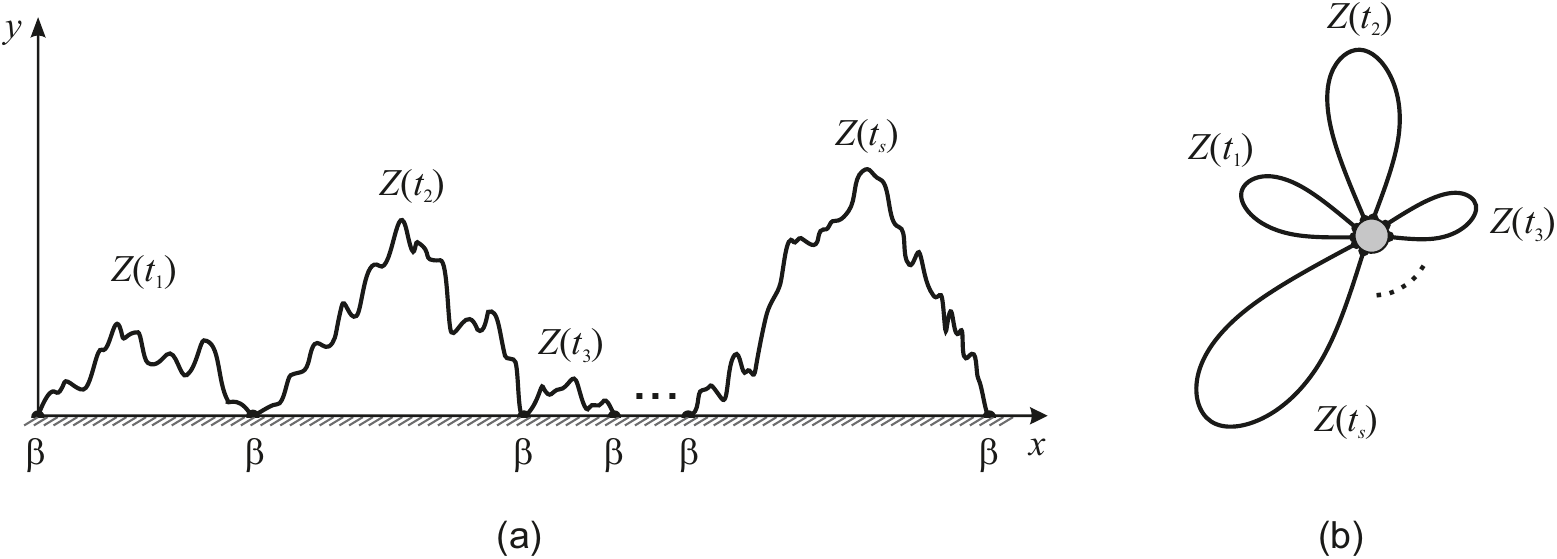}
\caption{Left: Representation of the adsorbed fractal polymer in terms of first returns to the point $x=0$; Right: schematic chamomile-like configuration of a fractal polymer interacting with a point-like potential well in a $D$-dimensional space (the drawing emphasizes the independence of loops).}
\label{fig:01}
\end{figure}

Consider the 1D problem of fractal polymer adsorption on a half-line $x\ge 0$ at a sticky surface located at $x=0$. Supposing that the polymer is grafted at its extremities at the point $x=0$, we can represent a polymer configuration as a consecutive set of $s$ ``bridges'' of lengths $t_1, t_2, t_3,...,t_s$. Within each bridge, the polymer does not touch the point $x=0$. So, the partition function $Z(t_j)$ of the polymer subchain of length $t_j$ ($j=1,...,s$) is the ``first return'' partition function, which can be written in the following generic form
\be
Z(t_j) \approx \frac{\lambda^{t_j}}{t_j^{\alpha}}
\label{eq:01}
\ee
where $\lambda$ accounts for the statistical weight of each monomer and is a non-universal quantity since it depends on the space dimensionality, lattice structure, polymer flexibility, etc. Besides, the critical exponent $\alpha= \alpha(H)$ is universal and characterizes the first return probability of a polymer with a given Hurst exponent $H$ on a half-line $x\ge 0$. Schematically, the particular polymer configuration is depicted in \fig{fig:01}: the panel (a) represents the path in the upper $(t,x)$-plane, while the panel (b) designates the typical chamomile-like structure of the path in a $D$-dimensional space interacting with a point-like potential well.

Denoting by $\beta$ the Boltzmann weight of a monomer located at the point $x=0$, we can represent the polymer partition function $Z_N(\beta)$ (where $N=t_1+t_2+t_3+...+t_s$) in the following form:
\be
Z_N(\beta) = \beta \sum_{s=1}^{\infty}\sum_{\{t_1+...+t_s=N\}} \prod_{j=1}^s (\beta Z(t_j)) = \beta \sum_{s=1}^{\infty}\; \sum_{t_1=1}^{\infty}...\sum_{t_s=1}^{\infty}\Delta(t_1+...+t_s-N) \prod_{j=1}^s (\beta Z(t_j))
\label{eq:02}
\ee
Using the integral representation of the $\Delta$-function
\be
\Delta(x) = \frac{1}{2\pi i}\oint \frac{d\xi}{\xi^{x+1}}=\begin{cases} 1 & x=0 \medskip \\ 0 & x\neq 0 \end{cases}
\label{eq:03}
\ee
with $x=N-(t_1+...+t_s)$, we get
\be
Z_N(\beta) = \frac{\beta\lambda^N}{2\pi i}\oint \frac{d\xi}{\xi^{N+1}} \sum_{s=1}^{\infty}\left(\beta\sum_{t=1}^{\infty}\frac{\xi^{t}}{t^{\alpha}}\right)^s = 
\frac{\beta\lambda^N}{2\pi i}\oint \frac{d\xi}{\xi^{N+1}} \frac{1}{1-\beta\, {\rm Li}_{\alpha}(\xi)}
\label{eq:04}
\ee
where ${\rm Li}_{\alpha}(\xi)$ is the polylogarithm function.

Since the value $\xi=1$ corresponds to $N\to\infty$, the divergence of the denominator in \eq{eq:04} at $\xi=1$  defines the phase transition point, $\beta_{tr}$, in the infinitely long chain:
\be
\beta_{tr} = {\rm Li}_{\alpha}^{-1}(1) = \zeta^{-1}(\alpha)
\label{eq:05}
\ee
where $\zeta(\alpha)$ is the Riemann $\zeta$-function. The order of the phase transition, $\gamma$, can be found by evaluating the first non-vanishing finite-size correction to the free energy, $F= -T\ln Z(\beta)$ of the system at large but finite $N$:
\be
F(\beta) = F(\beta_{tr}) + {\rm const}\,N |\beta-\beta_{tr}|^{\theta} \qquad (|\beta-\beta_{tr}|\ll 1)
\label{eq:06}
\ee
The transition order, $\gamma$, depends on the ``loop factor'' (i.e. on the first return probability to the point $x=0$) which is controlled by the critical exponent $\alpha$. Expanding ${\rm Li}_{\alpha}(\xi)$ in the vicinity of the point $\xi=1$ (note that $0<\xi\le 1$), we have
\be
{\rm Li}^{-1}_{\alpha}(\xi)\Big|_{\xi\nearrow 1} =\zeta^{-1}(\alpha) + \frac{|\Gamma(1-\alpha)|}{\zeta^2(\alpha)} |1-\xi|^{\alpha-1} 
\label{eq:07}
\ee
From \eq{eq:07} we can establish the relation between the deviations $|\xi-1|\ll 1$ and $|\beta-\beta_{tr}|\ll 1$:
\be
|\beta-\beta_{tr}| \approx \frac{|\Gamma(1-\alpha)|}{\zeta^2(\alpha)} |1-\xi|^{\alpha-1} \quad \Rightarrow \quad 
|1-\xi| \approx \frac{\zeta^2(\alpha)}{|\Gamma(\alpha-1)|} |\beta-\beta_{tr}|^{1/(\alpha-1)}
\label{eq:08}
\ee
Note that the transition point in a chain of a finite length ($0<\xi<1$) has always bigger $\beta$ than the transition point $\beta_{tr}={\rm Li}^{-1}(\alpha)$ in a chain of infinite length for any $\alpha>1$. According to \eq{eq:04}, the partition function in the localized phase at $N\gg 1$ is determined by the pole, $\xi_0$ of the function $\left(1-\beta\, {\rm Li}_{\alpha}(\xi)\right)^{-1}$ in the vicinity of the point $\xi=1$. Thus we can define the free energy $F(\beta) = -T\ln Z_N$ and consider the normalized free energy per one monomer, $f(\beta)=F(\beta)/(NT)$ at $N\gg 1$:
\be
Z_N(\beta)\Big|_{N\gg 1} \approx \beta\lambda^N \xi_0^N; \qquad f(\beta) = -\ln \lambda - \ln \xi_0
\label{eq:09}
\ee
The requested pole $\xi_0$ is given by \eq{eq:08}, i.e.
\be
\xi_0 = 1-\frac{\zeta^2(\alpha)}{|\Gamma(\alpha-1)|} |\beta-\beta_{tr}|^{1/(\alpha-1)}
\label{eq:10}
\ee
Eq. \eq{eq:10} provides the final expression for the dependence of the order of the phase transition $\gamma$ (see \eq{eq:06}) on the critical exponent $\alpha$ in the loop factor of the propagator \eq{eq:01}. Substituting \eq{eq:10} into \eq{eq:09} we get for $|\beta-\beta_{tr}|\ll 1$:
\be
f(\beta) = -\ln \lambda + \frac{\zeta^2(\alpha)}{|\Gamma(\alpha-1)|} |\beta-\beta_{tr}|^{\theta}, \qquad \theta = \frac{1}{\alpha-1} \quad (\alpha>1)
\label{eq:11}
\ee
The 2nd order phase transition ($\theta=2$) corresponds to $\alpha = \tfrac{3}{2}$. The exponent $\alpha=\frac{3}{2}$ is the standard critical exponent of: (i) the {\it first} return probability for a random walk on a half-line $x\ge 0$ with a sticky boundary at $x=0$, or (ii) the return probability in a three-dimensional space for a problem of a random walk adsorption at a point-like potential well.

The exponent $\alpha$ in \eq{eq:01} is related to the fractal dimension, $D_f$, of the polymer. Recall that $D_f$ is determined by the relation $\sqrt{\la R^2\ra} \sim N^{1/D_f}$ where $\la R^2 \ra$ is the mean-square distance of the trajectory with open ends, or the square of the gyration radius for closed chain. For the standard random walk $D_f=2$, for subdiffusive paths $D_f>2$, while for superdiffusive ones $0<D_f<2$. For weakly correlated fractal polymers the relation between $D_f$ and $\alpha$ can be established using the ``image method'' \cite{image}, which provides the following equation in the one-dimensional case:
\be
\alpha = D_f^{-1}+1
\label{eq:12}
\ee
It is noteworthy that the relation \eq{eq:12} fails for the fractal Brownian motion (fBm) due to strong long-distance correlations perturbing the path structure in the vicinity of the boundary \cite{metzler}. Typically, the first return exponent $\alpha$ for fBm is expressed in terms of the Hurst exponent $H$ via the Molchan's formula $\alpha = 2 - H$ (see \cite{molchan}). Taking into account that by definition $H = D_f^{-1}$, we get for fBm: $\alpha = 2-D_f^{-1}$. However, the case of fBm is beyond the scope of our consideration. 

Substituting \eq{eq:12} into \eq{eq:11} and applying the image method, we can establish a simple relation between the phase transition order $\theta\equiv\theta_{D_f}$ and the fractal dimension $D_f$ of the polymer:
\be
\theta_{D_f} = D_f
\label{eq:13}
\ee
From \eq{eq:13} we conclude that the adsorption transition of a polymer with the fractal dimensions $D_f=3$ and $D_f=1$ are, respectively, of 3rd ($\theta=\theta_3=3$) and 1st ($\theta=\theta_1=1$) orders. In Table \ref{tab:01} we summarize various exponents and relations between them.

\begin{table}
\begin{tabular}{ccc}
\hline \hline
$\alpha$ (first return)  & \hspace{1cm} $D_f$ (fractal dimension) & \hspace{1cm} $\theta$ (transition order) \\ \hline
$4/3$ & \hspace{1cm} 3  & \hspace{1cm} 3 \\ \hline
$3/2$ & \hspace{1cm} 2  & \hspace{1cm} 2 \\ \hline
2 & \hspace{1cm} 1  & \hspace{1cm} 1 \\ \hline \hline
\end{tabular}
\caption{Summary of critical exponents and relations between them for polymer adsorption at a potential well on a half-line.}
\label{tab:01}
\end{table}

\section{Unconventional phase transition in an inhomogeneous random walk hopping problem on a 1D lattice}
\label{sect:3}

In this Section we study fluctuations of the midpoint of the Brownian bridge (Bb) on a one-dimensional nonuniform lattice $k=1,2,..., K$ ($K\to \infty$) with a sticky boundary at $k=1$. We demonstrate that the fractal dimension of the Bb depends on the random walk hopping amplitude, $b_k$, where $k$ is the distance from the boundary. Introducing nonuniformity to $b_k$ along the lattice allows for a change in the scaling of the size of the Bb. Consequently, by applying the results from the previous section, one can modify the order of the adsorption transition at a sticky boundary.

We consider the following scaling of hopping rates, $b_k$:
\be
b_k= (K-k)^{\chi} = \begin{cases} 1  & \mbox{uniform amplitudes, $\chi = 0$} \medskip \\ 
\sqrt{K-k} & \mbox{descending amplitudes, $\chi=1/2$} \end{cases}
\label{eq:hopping}
\ee
where $K$ is the lattice size. The corresponding master equation for the partition function, $\Phi_N(k)$, of the $N$-step polymer which ends at a level $k=1,2,...,K$ can be written as follows
\be
\begin{cases}
\disp \Phi_{N+1}(k) =  b_{k-1}\Phi_N(k-1) +b_k\Phi_N(k+1) + \beta K^\chi \delta_{k,1} \Phi_N(k) & \mbox{for $k=2,3,..., K$} \medskip \\
\Phi_N(k=0) =\Phi_N(k=K+1)=0 \medskip \\
\Phi_N(k) = \delta_{N,0}\,\delta_{k,1}
\end{cases}
\ee
where $\beta=e^U$ is the weight of potential well of depth $U$ at the boundary site $k=1$ and the Kronecker $\delta$-function is equal to 1 for $k=1$ and 0 otherwise. One can straightforwardly prove that the partition function of the Brownian bridge, $\Phi_N(k=0)$, can be written in two different ways, giving the same result:
\be
\Phi_N(k=1) = \left<{\bf v}_{in}|T^N|{\bf v}_{out}\right>
\ee
where
\be
T = \left(\begin{array}{cccccc}
\beta K^\chi & (K-1)^{\chi} & 0 & 0 & 0 & \dots  \\
(K-1)^{\chi} & 0 & (K-2)^{\chi} & 0 & 0 & \\
0 & (K-2)^{\chi} & 0 & (K-3)^{\chi} & 0 & \\
0 & 0 & (K-3)^{\chi} & 0 & (K-4)^{\chi} & \\
0 & 0 & 0 & (K-4)^{\chi} & 0 & \\
\vdots & & & & & \ddots
\end{array} \right), \quad 
{\bf v}_{in}={\bf v}_{out}^{\top}=
\left(\begin{array}{c} 1 \\ 0 \\ 0 \\ 0 \\ 0 \\ \vdots \end{array} \right)
\ee
or
\be
\Phi_N(k=1) = \left<{\bf v}_{in}|(T')^N|{\bf v}_{out}\right>
\ee
where
\be
T' = \left(\begin{array}{cccccc}
\beta K^\chi \; & 1\; & 0\; & 0\; & 0\; & \dots  \\
(K-1)^{2\chi}\; & 0\; & 1\; & 0\; & 0\; & \\
0\; & (K-2)^{2\chi}\; & 0\; & 1\; & 0\; & \\
0\; & 0\; & (K-3)^{2\chi}\; & 0\; & 1\; & \\
0\; & 0\; & 0\; & (K-4)^{2\chi}\; & 0\; & \\
\vdots & & & & & \ddots
\end{array}\right), \quad 
{\bf v}_{in}={\bf v}_{out}^{\top}=
\left(\begin{array}{c} 1 \\ 0 \\ 0 \\ 0 \\ 0 \\ \vdots \end{array} \right)
\ee
It was argued in \cite{gorsky2018statistical,valov2021equilibrium} that $T'$ at $\chi=1/2$ can be interpreted as a transfer matrix of a path counting problem on a finite tree of $K$ generations with a linearly descending vertex degree. In these works, it was also shown that such a scaling of vertex degree corresponds to the special mean-field representation of the Gaussian matrix ensemble \cite{dumitriu2002matrix}. Recently, the tridiagonal representation of matrices in the 1D hopping problem has been identified with the construction of the Krylov basis in matrix models \cite{balasubramanian2023tridiagonalizing}.

Here, we remain in the same paradigm as in the model I considered in the previous Section: the very presence of the transition is controlled by the depth of the potential well, $U$, at the boundary (i.e., by the ``stickiness'' $\beta$ ), while the order of the phase transition depends on the scaling exponent $\chi$ of the hopping amplitude. The statistics of a Bb on a 1D lattice is determined by the conditional probability $Q_N(k,n|K)$ to find the $n$'th step of a Bb at the distance $k$ from the boundary:
\be
Q_N(k,n|K)=\frac{P_n(k)P_{N-n}(k)}{\sum\limits_{k=1}^K P_n(k)P_{N-n}(k)}=\frac{1}{\cal N}[T^n]_{k,1}[T^{N-n}]_{k,1}
\label{eq22}
\ee
where $P_N(k)\propto \Phi_N(k)$ is the probability to find the end of the open path of length $N$ at the distance $k$, $[T^N]_{k,1}$ is the element $(k,1)$ of the matrix $T^N$, and ${\cal N}$ is the normalization constant for the conditional distribution $Q_N(k,n|K)$. In what follows without the loss of generality and for simplicity, we take $n=N/2$, i.e., we consider the distribution of the Brownian bridge midpoint. 

\begin{figure}[ht]
\includegraphics[width=0.6\textwidth]{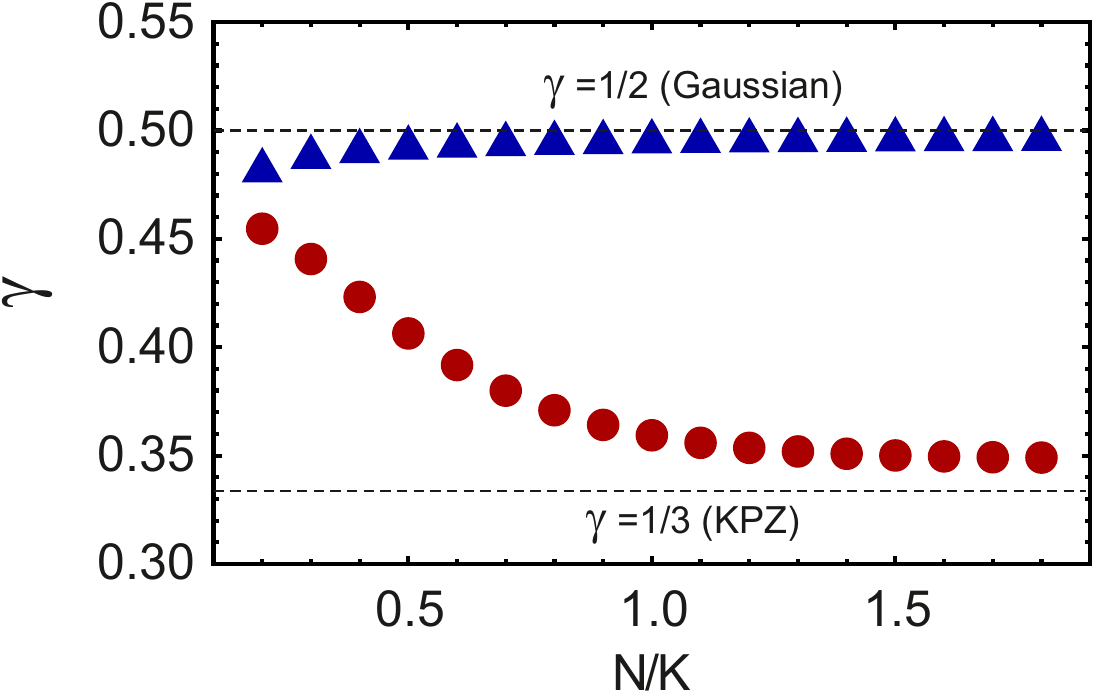}
\caption{Limiting behavior of critical exponent $\gamma$  of Brownian bridge in a double-scaling regime ($N\to\infty$, $K\to\infty$, $N/K={\rm const}$): Blue triangles -- for the uniform hopping rates ($\chi=0$), red circles -- for the descending hopping rates ($\chi=1/2$).}
\label{fig:05}
\end{figure}

Equation \eq{eq22} permits us to compute the scaling of a typical span $\Delta(N) \sim N^{\gamma}$ of a Brownian bridge on a 1D lattice with {\it descending} hopping amplitudes in the doubly scaling limit $N\to\infty$, $K\to\infty$, $N/K={\rm const}$ and find the critical exponent $\gamma$, which coincides with the Kardar-Parisi-Zhang (KPZ) growth exponent, $\gamma_{KPZ}=1/3$. To the contrary, the critical exponent $\gamma$ for the Bb on a lattice with {\it uniform} transition rates has been repeatedly computed (see, for example, \cite{khokhlov}) where it has been shown that such a Bb has a span $\Delta$ controlled by the exponent $\gamma_{Gauss}=1/2$. The plots demonstrating the saturation of the exponent $\gamma$ for the lattices with uniform and descending amplitudes are shown in \fig{fig:05}. 

\begin{figure}[ht]
\includegraphics[width=0.86\textwidth]{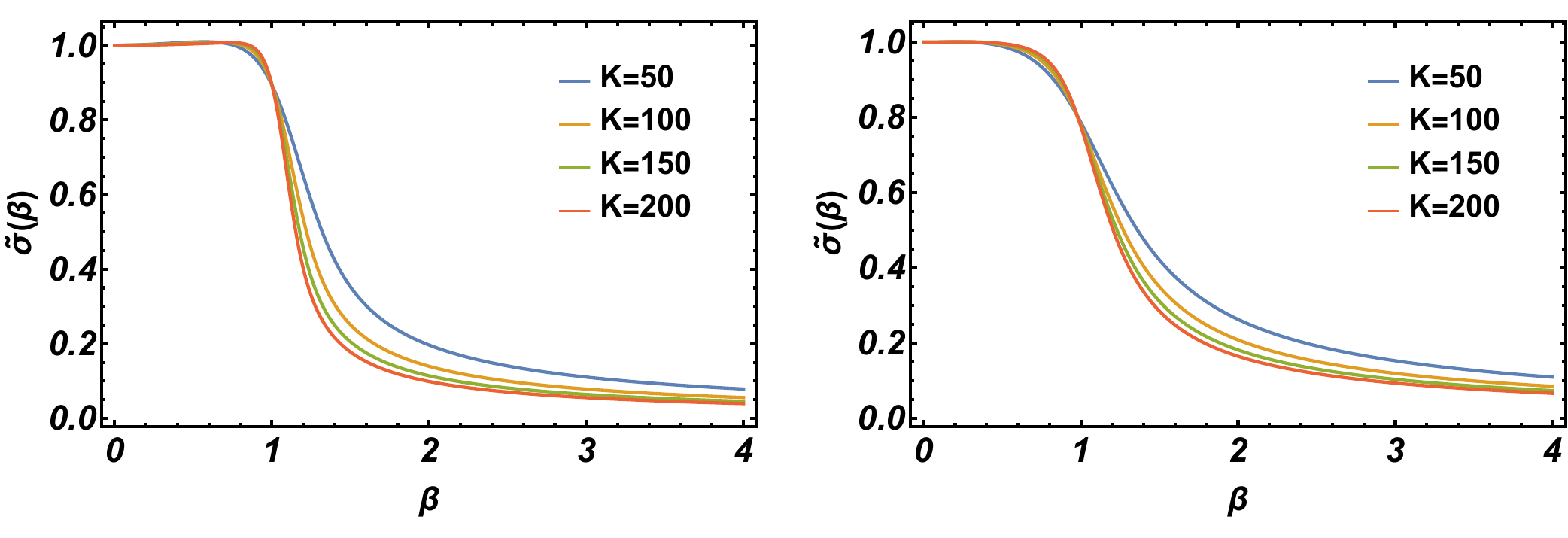}
\caption{Dependence of the rescaled variance, $\tilde{\sigma}$ on $\beta$ for a few values of $K$: (a) Uniform hopping amplitudes ($\chi=0$); (b) Descending hopping amplitudes ($\chi=1/2$).}
\label{fig:06}
\end{figure}

To determine numerically the order of the localization transition of a Bb on the 1D lattice with a sticky boundary, we proceed as follows. First, we define the dependence of the rescaled variance, $\tilde{\sigma}(\beta)=\sigma(\beta)/\sigma(\beta=0)$, of the Brownian bridge midpoint on the stickiness $\beta$ for two different kinds of hopping amplitudes (uniform and descending) and for various lattice sizes $K$. The corresponding plots are shown in \fig{fig:06}a,b.

\begin{figure}[ht]
\centering
\includegraphics[width=0.9\textwidth]{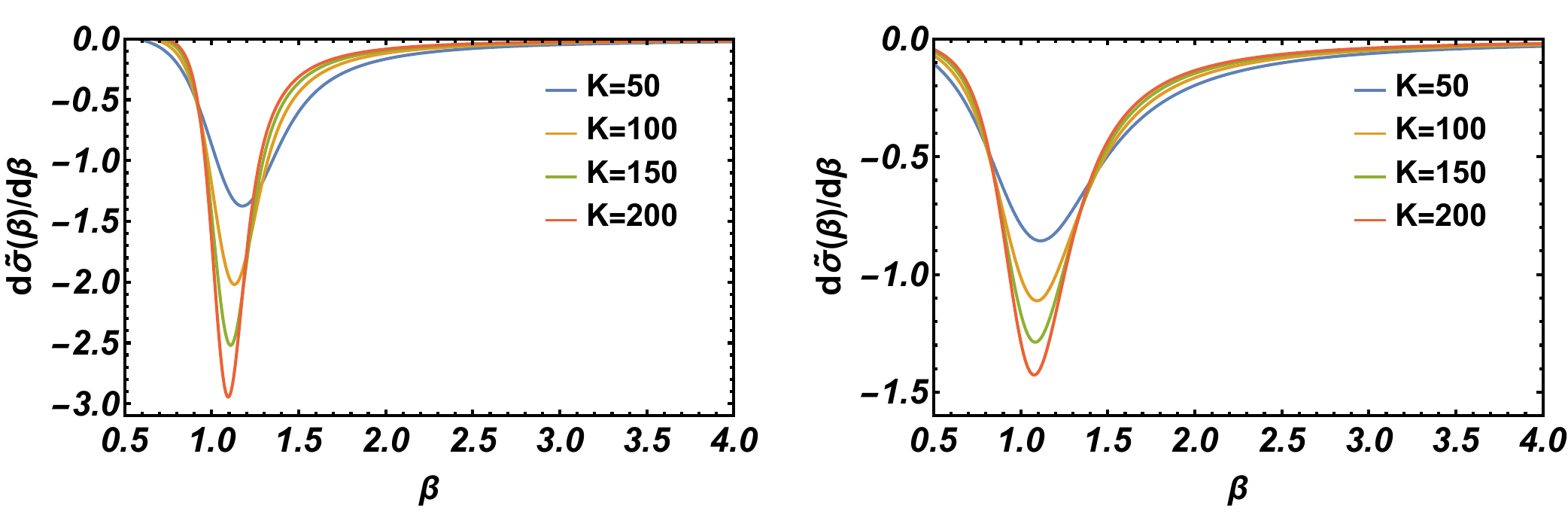}
\caption{Plot of the function $\tilde{\sigma}'(\beta)$: (a) Uniform hopping amplitudes ($\chi=0$); (b) Descending hopping amplitudes ($\chi=1/2$).}
\label{fig:07}
\end{figure}

Next, we numerically compute the derivative $\tilde{\sigma}'(\beta)\equiv \frac{d\tilde{\sigma}(\beta)}{d\beta}$ and associate the transition width, $\Delta$, with the width of the function $\tilde{\sigma}'(\beta)$ at the level $\min(\sigma(\beta))/\sqrt{2}$ for every $K$ -- see \fig{fig:07}a,b for uniform and descending hopping amplitudes.

\begin{figure}[ht]
\centering
\includegraphics[width=0.6\textwidth]{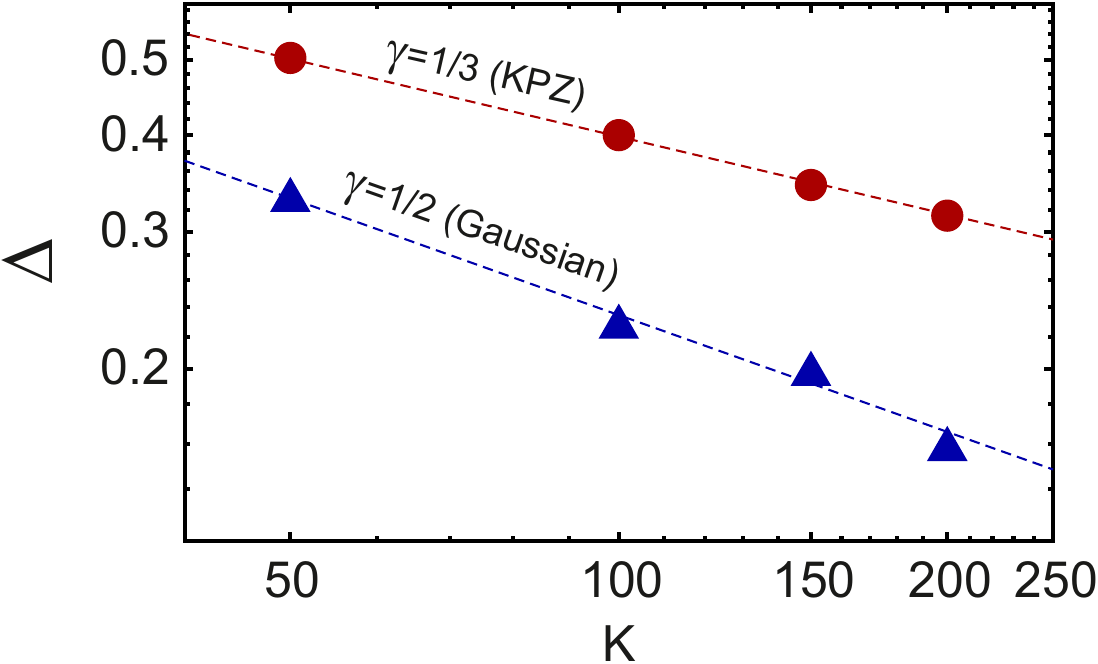}
\caption{The finite-size scaling (in a log-log scale) of the width of adsorption transition as a function of a size of a tree: the Gaussian exponent $\gamma=1/2$ for uniform amplitudes (blue triangles); the KPZ exponent $\gamma=1/3$ for descending amplitudes (red circles).}
\label{fig:08}
\end{figure}

It is well known \cite{binder} that the order of the phase transition, $\theta$, can be extracted from the finite-size dependence of the transition width, $\Delta$ on $K$. Namely, if $\Delta$ shrinks with $K$ as $\Delta\sim K^{-1/\theta}$, then the transition order at $K\to\infty$ is $\theta$. The dependence of the width $\Delta$ on the lattice size, $K$, together with the power-law approximation $a K^{-1/\theta}$, are plotted in \fig{fig:08} in doubly-logarithmic coordinates. 

For both rates $b_k$, uniform ($\chi=0$) and descending ($\chi=1/2$), the localization transition width, $\Delta$, behaves as $\Delta \propto K^{-1/\theta}$ where $\theta=D_f$ and $D_f$ is the fractal dimension of Brownian bridge on the corresponding lattice. From plots shown in \fig{fig:08} we conclude that $\theta_{Gauss}=D_f=2$ (the 2nd order phase transition) for a uniform hopping amplitudes ($\chi=0$), and $\theta_{KPZ}=D_f=3$ (the 3rd order phase transition) for a descending hopping amplitudes with $\chi=1/2$.

\section{Discussion}
\label{sect:4}

In this note, we discuss random walks on the semi-infinite line with a sticky boundary, exploring various scaling laws of path dynamics. The main attention was addressed to the question how the order of the localization phase transition at a critical value of the boundary stickiness depends on scaling details of the random walk in bulk. In Section \ref{sect:2} it was demonstrated that the order of the phase (adsorption) transition can be changed by tuning the scaling exponent of the first return probability. In Section \ref{sect:3} we discussed the localization of the closed inhomogeneous random walk (the inhomogeneous Brownian bridge) on a one-dimensional lattice at a sticky boundary. We have shown that by changing the scaling exponent $\chi$ of local hopping amplitudes from $\chi=0$ to $\chi = 1/2$ as it is defined in \eq{eq:hopping}, the 2nd order localization transition gets transformed into the 3rd order one.

A similar inhomogeneous hopping problem on a one-dimensional lattice can be viewed as motion in the Krylov space. We will explore related aspects of criticality in an upcoming publication, where we will also address nonuniform lattices with scaling of hopping amplitudes $\chi$ different from 0 and 1/2 \cite{gnv}.

\begin{acknowledgments}
We are grateful to M. Tamm for valuable discussions on polymer adsorption problem and to R. Metzler and M. Dolgushev for elucidating questions concerning the interplay between return probability and span of a fractal polymer chain.
\end{acknowledgments}

\bibliography{ref.bib}

\end{document}